\documentclass[
aps,
prc,
showpacs,
superscriptaddress,
nofootinbib,
floatfix]
{revtex4-1}
\usepackage{graphicx,
longtable,
slashed}

\begin{document}
%
\title{Possibility of measuring the CP Majorana phases in $0\nu\beta\beta$-decay}
%
%
\author{        F.~\v{S}imkovic}
\affiliation{\it  Comenius University, Mlynsk\'a dolina F1, SK--842 48
Bratislava, Slovakia and IEAP CTU, 128--00 Prague, Czech Republic}
\affiliation{	Bogoliubov Laboratory of Theoretical Physics, JINR, 
				141980 Dubna, Moscow region, Russia}
				
\author{        S.M.~Bilenky}
\affiliation{	Bogoliubov Laboratory of Theoretical Physics, JINR, 
				141980 Dubna, Moscow region, Russia}
\affiliation{   TRIUMF 4004 Wesbrook Mall, Vancouver BC, V6T 2A3 Canada}

\author{        Amand~Faessler}
\affiliation{   Institute of Theoretical Physics,
				University of Tuebingen, 
               72076 Tuebingen, Germany}

\author{        Th.~Gutsche}
\affiliation{   Institute of Theoretical Physics,
				University of Tuebingen, 
               72076 Tuebingen, Germany}
\begin{abstract}
In view of recent measurements of the mixing angle $\theta_{13}$ we investigate
the possibility to  determine the difference of two CP Majorana phases 
of the neutrino mixing matrix from the study of neutrinoless double-beta decay.
We show that if cosmological measurements will reach the sensitivity of 0.1 eV 
for the sum of neutrino masses, i.e. the mass value of the lightest neutrino 
will be strongly constrained, the long-baseline neutrino oscillation experiments
will determine inverted hierarchy of neutrino masses and 
if neutrinoless double-beta  decay will be observed, this determination
might be possible. The required experimental accuracies and the uncertainties 
in the calculated nuclear matrix elements of the process are discussed in this context.
\end{abstract}
\medskip
\maketitle

\section{Introduction \label{SecI}}

The discovery of neutrino oscillations in the SuperKamiokande
(atmospheric neutrinos) \cite{SUPERKAMIOKANDE}, SNO (solar neutrinos)
\cite{SOLAROSC}, KamLAND (reactor neutrinos) \cite{KAMLAND}, MINOS \cite{Minos}
(accelerator neutrinos) and other neutrino experiments gives us
compelling evidence that neutrinos possess small masses and
flavor neutrino fields are mixed.

Neutrino flavor states $|\nu_\alpha>$ ($\alpha = e,~\mu, ~\tau$) are
connected to the states of neutrinos with  masses $m_j$
($|\nu_j\rangle$)  by the following standard mixing relation
\begin{eqnarray}
|\nu_\alpha\rangle = \sum^{3}_{j=1} U^{*}_{\alpha j} |\nu_j\rangle
~~~~~(\alpha = e,~\mu, \tau ),
\end{eqnarray}
where $U$ is the $3\times3$ Pontecorvo-Maki-Nakagawa-Sakata
(PMNS) unitary mixing matrix. If the massive neutrinos are Dirac
(Majorana) particles the PMNS matrix  contains one
(three) $CP$ phases.

The problem to determine the CP phases is one of the
major challenges in today's neutrino physics. Some information about
lepton phases could help to solve the problem of the baryon asymmetry
of the Universe \cite{fuku86}. The discovery in the Daya Bay
\cite{dayabay}, RENO \cite{reno12}, T2K \cite{T2K} and Double Chooz
\cite{doublech}  experiments of a relatively large mixing
angle $\theta_{13}$ opened a possibility of measuring
the Dirac phase $\delta$ in long baseline accelerator experiments. 
The Majorana $CP$ phases can only be determined through the observation of 
neutrinoless double $\beta$-decay ($0\nu\beta\beta$-decay)   \cite{BHP80,SV80,langan}.

The $CP$ phases enter into the effective Majorana mass
\cite{vissani99,barger02,gouvea03,pascol02,pascol06} defined as
\begin{equation}
m_{\beta\beta} = \sum_j  U^2_{ej} m_j.
\end{equation}
This quantity depends also on the neutrino oscillation parameters
$\theta_{12}$, $\theta_{13}$, $\Delta m^2_{\mbox{\tiny{SUN}}}$
$\Delta m^2_{\mbox{\tiny{ATM}}}$, the lightest neutrino mass and the type
of the neutrino mass spectrum (normal or inverted). 

The effective Majorana mass $m_{\beta\beta}$ can be determined in the
neutrinoless double beta decay of even-even nuclei  \cite{feruglio,AEE08,VES12}
\begin{equation}
(A,Z) \rightarrow (A,Z+2) + 2 e^-
\end{equation}
by relating the $0\nu\beta\beta$-decay half-life to
$|m_{\beta\beta}|$ using calculated nuclear matrix elements (NMEs).

The goal of this paper is to discuss a possibility for determining
the Majorana $CP$ phases from data of
$0\nu\beta\beta$ experiments of the next generation assuming that the
$0\nu\beta\beta$-decay will be observed. The
problem of the experimental accuracies and the theoretical uncertainties of the
calculated NMEs will also be addressed.

\section{Neutrino oscillations and effective Majorana  mass}

It was proved in experiments with atmospheric, solar,
reactor and accelerator neutrinos that flavor neutrinos oscillate
from one flavor (electron-, muon-, and tau-) to another due to
neutrino mixing and non-zero neutrino mass-squared  differences. All
existing neutrino oscillation data (with the exception of the LSND \cite{lsnd},
MiniBooNE \cite{MiniBOONE}, short baseline reactor \cite{RNA11} and Gallium  \cite{galanom}) anomalies)
are perfectly described by the minimal scheme of three-neutrino mixing.

In the case of Dirac neutrinos the unitary $3\times 3$ PMNS neutrino mixing matrix can be
parameterized  as follows
 \begin{equation}
 U = \left(
\begin{array}{lll}
 c_{12} c_{13} & c_{13} s_{12} & e^{-i \delta } s_{13} \\
 -c_{23} s_{12}-e^{i \delta } c_{12} s_{13} s_{23} & c_{12}
   c_{23}-e^{i \delta } s_{12} s_{13} s_{23} & c_{13} s_{23} \\
 s_{12} s_{23}-e^{i \delta } c_{12} c_{23} s_{13} & -e^{i \delta }
   c_{23} s_{12} s_{13}-c_{12} s_{23} & c_{13} c_{23}
\end{array}
\right)
\label{pmns}
 \end{equation}
where  $c_{ij}\equiv \cos{\theta_{ij}}$, $s_{ij}\equiv
\sin{\theta_{ij}}$. $\theta_{12}$, $\theta_{13}$ and $\theta_{23}$
are mixing angles and $\delta$ is the CP phase. If
neutrinos are Majorana particles the matrix $U$ in Eq. (\ref{pmns})
is multiplied by a diagonal phase matrix 
$P = diag(e^{i (\alpha_1/2-\delta)}, e^{i (\alpha_2/2-\delta)}, 1)$,
which contains  two additional CP phases $\alpha_1$ and
$\alpha_2$.

With the discovery of neutrino oscillations we know:
    \begin{itemize}
    \item The values of the large mixing angles $\theta_{12}$ and $\theta_{23}$.
The value of the relatively small angle $\theta_{13}$ recently measured in
the Double Chooz \cite{doublech},  the Daya Bay \cite{dayabay} and
              RENO \cite{reno12} reactor neutrino experiments.
\item The  solar and atmospheric mass-squared differences\footnote{We use the following
definition $\Delta m^2_{ij}=m^{2}_{j}-m^{2}_{i}$ } $\Delta
m^2_{\mbox{\tiny{SUN}}}=\Delta m^2_{12}$ and $\Delta m
^2_{\mbox{\tiny{ATM}}}=\Delta m^2_{23}$ (Normal  spectrum),  $\Delta
m^2_{\mbox{\tiny{ATM}}}=-\Delta m^2_{13}$ (Inverted spectrum).
\end{itemize}
We do not know the value of the lightest neutrino mass, the $CP$ phases and
the character of the neutrino mass spectrum (normal or inverted).

From the data of the MINOS experiment \cite{Minos} it was found that
$\Delta m^2_{\mbox{\tiny{ATM}}}=(2.43 \pm 0.13) \times
10^{-3}~\mathrm{eV}^{2}$. From the analysis of the KamLAND and solar
data it was obtained that $\tan^2{\theta_{12}}=0.452_{-0.033}^{+0.035}$
\cite{KAMLAND}. From the global fit to all data
it was inferred that \cite{SCHWETZ} $\Delta m^2_{\mbox{\tiny{SUN}}}= (7.65^{+0.13}_{-0.20})
\times 10^{-5}~\mathrm{eV}^{2}$ and
$\sin^2{\theta_{23}}=0.50_{-0.06}^{+0.07}$. Finally from the analysis
of the Daya Bay \cite{dayabay} and RENO data\cite{reno12}
one obtains $\rm{\sin}^2 2 \theta_{13} = 0.092 \pm
0.016~(stat)~\pm 0.005~(syst)$ and $\rm{\sin}^2 2 \theta_{13} = 0.103
\pm 0.013~(stat)~\pm 0.011~(syst)$, respectively.

The effective Majorana  mass is given by
\begin{eqnarray}
|m_{\beta\beta}| = |c^2_{12}c^2_{13} e^{i\alpha_1} m_1 + s^2_{12}
c^2_{13} e^{i\alpha_2} m_2 + s^2_{13} m_3| \label{efmas}
\end{eqnarray}
or by the full expression
\begin{eqnarray}
|m_{\beta\beta}|^2 &=&
c^4_{12} c^4_{13} m^2_1 +  s^4_{12}  c^4_{13} m^2_2 + s^4_{13} m^2_3 \nonumber\\
&& + 2 c^2_{12} s^2_{12} c^4_{13} m_1 m_2 \cos{(\alpha_1 - \alpha_2)} \nonumber\\
&& + 2 c^2_{12} c^2_{13} s^2_{13} m_1 m_3 \cos{\alpha_1}
   + 2 s^2_{12} c^2_{13} s^2_{13} m_2 m_3 \cos{\alpha_2}.
\label{eq.6}
\end{eqnarray}
From this equation it simply follows that the effective Majorana mass
depends on the character of the neutrino mass spectrum and three
unknown parameters: the lightest neutrino mass and the two $CP$
phases. We note that for two set of phases $\alpha_1, \alpha_2$ and 
$(2\pi-\alpha_1)$, $(2\pi-\alpha_2)$ the same value of $|m_{\beta\beta}|$ 
is reproduced.

\begin{figure}[!t]
\includegraphics[scale=.52]{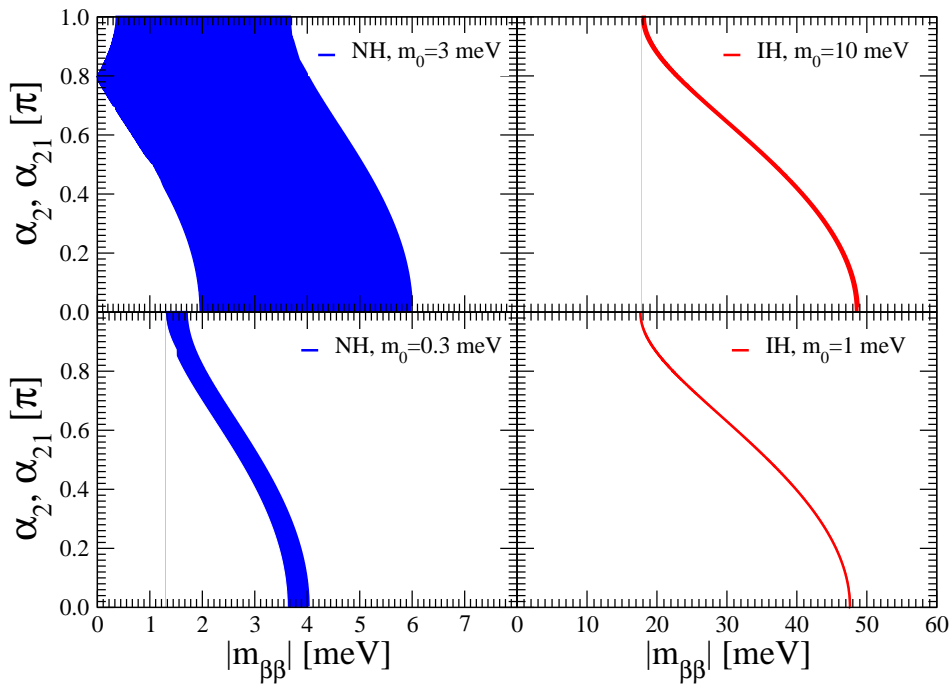}
\vspace*{+1.0cm}
\caption{(Color online) The CP phase $\alpha_2$ (left panels)
and difference of phases $\alpha_{21}$ (right panels) are plotted, respectively,
as function of the absolute value of the effective majorana mass  for
the IH and NH of neutrino masses and the chosen value of $m_0$.
\label{fig.1}}
\end{figure}

In the three-neutrino case two mass spectra are currently possible:
\begin{itemize}
\item { Normal Spectrum (NS)}: $m_{1} < m_{2} < m_{3}$:
~~$\Delta m^{2}_{12}\ll \Delta m^{2}_{23}$.
In this case
$$m_2=\sqrt{{ \Delta m^2_{\mbox{\tiny{SUN}}}}+{ m_0^2}}~,
~m_3=\sqrt{{\Delta m^2_{\mbox{\tiny{ATM}}}}+\Delta
m^2_{\mbox{\tiny{SUN}}}+{ m_0^2}}$$ with $m_0= m_1$.
\item { Inverted Spectrum (IS)}, $m_{3} < m_{1} < m_{2}$:~~
$\Delta m^{2}_{12}\ll |\Delta m^{2}_{13}|$.
We have
$$ m_1 = \sqrt{ \Delta m^2_{\mbox{\tiny{ATM}}}+{ m_0^2}},~~~
m_2=\sqrt{{ \Delta m^2_{\mbox{\tiny{ATM}}}+\Delta
m^2_{\mbox{\tiny{SUN}}}}+{ m_0^2}}$$ with $m_0=m_3$.
\end{itemize}
For both cases $m_{0}=m_{1}(m_{3})$ is the lightest neutrino mass.

For the two neutrino mass hierarchies we can set constraints on the effective Majorana
mass:
\begin{enumerate}
\item Normal Hierarchy (NH): $m_1\ll m_2 \ll m_3$:\\
In this case for the neutrino masses we have
$$m_1 \ll \sqrt{\Delta m^2_{\mbox{\tiny{SUN}}}}.~~~m_2 \simeq \sqrt{\Delta m^2_{\mbox{\tiny{SUN}}}},
~~~m_3 \simeq  \sqrt{\Delta m^2_{\mbox{\tiny{ATM}}}}. $$
Neglecting the negligibly small contribution of $m_1$ we find
\begin{equation}
\cos{\alpha_2} \simeq \frac{|m_{\beta\beta}|^2 - s^4_{12}  c^4_{13} \Delta m^2_{\mbox{\tiny{SUN}}}
- s^4_{13} \Delta m^2_{\mbox{\tiny{ATM}}}}
{2 s^2_{12} c^2_{13} s^2_{13} \sqrt{\Delta m^2_{\mbox{\tiny{SUN}}} \Delta m^2_{\mbox{\tiny{ATM}}}}}.
\end{equation}
For the effective Majorana mass we then have the following range of values
\begin{equation}
|s^2_{12}  c^2_{13} \sqrt{\Delta m^2_{\mbox{\tiny{SUN}}}}
- s^2_{13} \sqrt{\Delta m^2_{\mbox{\tiny{ATM}}}}|
\le |m_{\beta\beta}| \le
s^2_{12}  c^2_{13} \sqrt{\Delta m^2_{\mbox{\tiny{SUN}}}}
+ s^2_{13} \sqrt{\Delta m^2_{\mbox{\tiny{ATM}}}}.
\end{equation}
Using the best-fit values of the mass squared differences and the mixing
angles we find
$$1.5~ \mathrm{meV} \le |m_{\beta\beta}| \le 3.8 ~\mathrm{meV }~.$$.
\item  Inverted Hierarchy (IH): $m_3 \ll m_1 < m_2$:\\
In the IH scenario  $m_3 \ll \sqrt{\Delta m^2_{\mbox{\tiny{ATM}}}}$ and
$m_1\simeq m_2 \simeq \sqrt{\Delta m^2_{\mbox{\tiny{ATM}}}}$.  We
find
\begin{equation}
\cos{\alpha_{12}} = \frac{|m_{\beta\beta}|^2 -
c^4_{13}(1-2s^2_{12}c^2_{12}) \Delta m^2_{\mbox{\tiny{ATM}}}} {2
c^2_{12} s^2_{12} c^4_{13} \Delta m^2_{\mbox{\tiny{ATM}}}}
\end{equation}
where $\alpha_{12} = \alpha_1 - \alpha_2$. For the absolute
value of the effective Majorana mass we have
\begin{equation}
|\cos 2\theta_{12}| c^2_{13} \sqrt{\Delta m^2_{\mbox{\tiny{ATM}}}}
\le |m_{\beta\beta}| \le c^2_{13} \sqrt{\Delta m^2_{\mbox{\tiny{ATM}}}}.
\end{equation}
Using the best-fit values of the parameters  we find the following
range for $|m_{\beta\beta}|$ in the case of the IH:

$$18~\mathrm{meV} \le |m_{\beta\beta}| \le 48~\mathrm{meV}~.$$

\end{enumerate}

The absolute value of the neutrino mass can be determined from
a precise measurement of the end-point part of the $\beta$-spectrum
of the tritium \cite{otten} and other $\beta$-decay
measurements \cite{mare}. Cosmological observations allow to infer
the sum of the neutrino masses
 \begin{equation}
m_{\mbox{\tiny cosmo}}
  =  \sum^{3}_k   m_k.
 \end{equation}
For inverted and normal hierarchy of neutrino masses there is a minimal value 
of $m_{\mbox{\tiny cosmo}}$ allowed by  the oscillation data as follows:
\begin{eqnarray}
m_{\mbox{\tiny cosmo}}  &\simeq& 2\sqrt{ \Delta m^{2}_{ATM}}\simeq 105~\mathrm{meV}~~~(\text{IH})
\nonumber\\
&\simeq& \sqrt{ \Delta m^{2}_{ATM}}\simeq ~62~\mathrm{meV}~~~(\text{NH}).
\end{eqnarray}

The current limits  on $m_{\mbox{\tiny cosmo}}$ depend on the type of
observations included in the fit \cite{Abarajan11}. The CMB primordial  gives  $\leq$ 1.3
eV, CMB+distance $\leq$  0.58 eV, galaxy distribution and lensing of
galaxies $\leq$ 0.6 eV. On the other hand  the largest photometric
red shift survey yields $\leq$ 0.28 eV \cite{ThAbdaLah10}. It is expected that
future  cosmological observables will provide precise constraints on the sum of
neutrino masses $m_{\mbox{\tiny cosmo}}$ \cite{cosmomass}. These constraints will
be such that they are even sensitive
to the minimal values of 0.105 eV and 0.062 eV allowed by the oscillation data
for the IH and NH, respectively (see, e.g., the recent summary \cite{Abarajan11}).
In the case of the IH and for the lowest value of $m_{\mbox{\tiny cosmo}}$
the value of the lightest neutrino mass $m_0$ can be  restricted to values
below a value of about 10 meV depending on the accuracy of the cosmological measurement.
We note that the neutrino mass hierarchy can be probed with accelerator based neutrino 
oscillation experiments through earth matter effects \cite{hierosc1,hierosc2}. However, 
from neutrino oscillation experiments alone one cannot determine the absolute 
neutrino mass scale or even constrain the mass of the lightest neutrino unlike 
for the case of cosmological measurements.

In Fig. \ref{fig.1}, by exploiting Eq. (\ref{eq.6}), the Majorana CP phase $\alpha_2$
(or difference of phases $\alpha_{21}=\alpha_2-\alpha_1$) is plotted as function of
the absolute value of the effective Majorana mass for chosen values of $m_0$ and
by assuming  the NH (IH) of neutrino masses. The second phase $\alpha_1$ is considered
to be arbitrary.
The results strongly depend on the value of $m_0$
and the type of neutrino mass  hierarchy, normal or inverted.
We find that when $m_0$ lies within a range of 0 to 10 meV and when
the IH is considered, the phase difference $\alpha_{21}$ depends only weakly on
$|m_{\beta\beta}|$. This is due to the fact that the third term on the right hand side
of Eq.  (\ref{efmas}) is small in comparison with the first two terms.
A different situation occurs in the case of the NH.
The results depend strongly
on $m_0$ in the considered range of (0-10) meV. There is practically no chance
to determine a value for $m_0$ by any laboratory or cosmological measurement,
if it is lower than a few meV. Thus, by measuring $0\nu\beta\beta$-decay it will not
be possible to obtain model independent information on the value of at least
one of the three CP Majorana phases when there is a normal hierarchy of
neutrino masses. One can rely only on those particle physics models which
allow to predict all three neutrino masses. In these cases values for one of
the CP Majorana phases could be obtained when considering the second phase to
be arbitrary and by observing the $0\nu\beta\beta$-decay. From Fig. \ref{fig.1}
it follows that if $m_0$ is about 3 meV this possibility is also very much limited.
It is interesting to note that for this value of $m_0$ the minimal
value of $|m_{\beta\beta}|$ does not appear for the case of CP-conservation
(see left upper panel of Fig. \ref{fig.1}) but for $\alpha_2 \simeq 0.79\pi$.

\begin{figure}[!t]
\includegraphics[scale=.52]{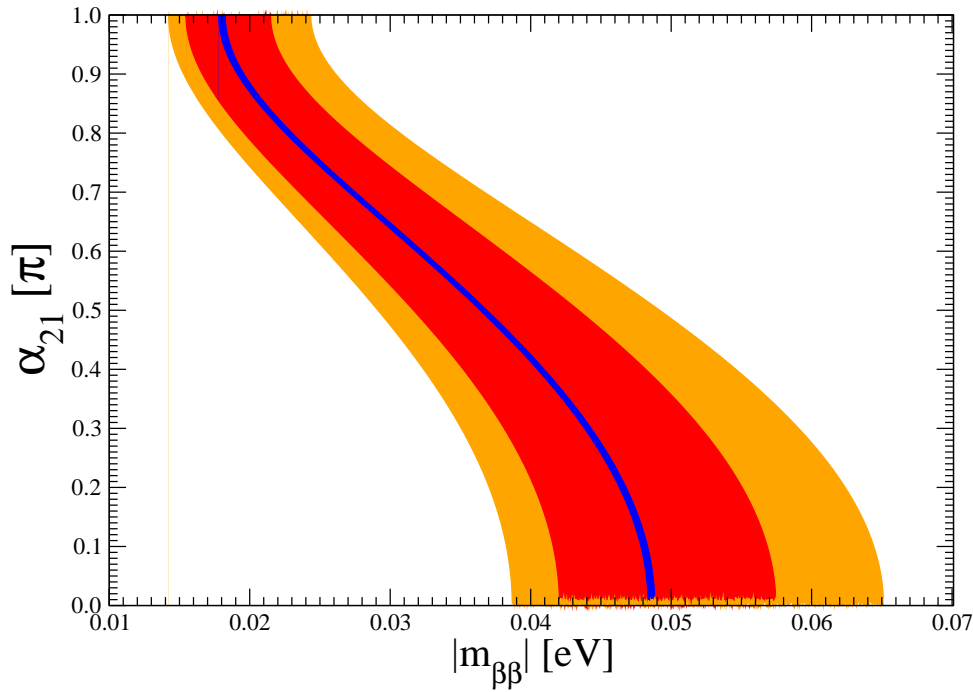}
\vspace*{+1.0cm}
\caption{(Color online)
The difference of CP phases $\alpha_{21}=\alpha_2-\alpha_1$ plotted
as function of $|m_{\beta\beta}|^{obs}$ for the inverted hierarchy of neutrino masses.
The current experimental errors of neutrino mixing parameters and mass
squared differences are taken into account; $m_0$ is taken to be in
the range of 0 to 10 meV. The blue, red and orange regions correspond
to $|m_{\beta\beta}|^{obs}$ with errors ($\sigma_{\beta\beta}/|m_{\beta\beta}|^{obs}$)
of 0\%, 15\% and 25\%, respectively.
\label{fig.2}}
\end{figure}

\section{Effective Majorana mass and the $0\nu\beta\beta$-decay}

Assuming that the $0\nu\beta\beta$-decay is driven by the Majorana neutrino mass mechanism, 
we have for the effective Majorana mass $m_{\beta \beta}$
\begin{eqnarray}
|m_{\beta\beta}| = \frac{m_e}
{ \sqrt{T^{0\nu}_{1/2} G^{0\nu}(Q_{\beta\beta},Z)} g^2_A |{M'}^{0\nu}| }.
\end{eqnarray}
Here, $T^{0\nu}_{1/2}$, $G_{0\nu}(Q_{\beta\beta},Z)$, $g_A$ and ${M'}^{0\nu}$ are,
respectively, the half-life of the $0\nu\beta\beta$-decay, the known phase-space factor,
the unquenched axial-vector coupling constant and the nuclear matrix element, 
which depends on nuclear structure. Recently, a complete and improved calculation 
of phase-space factors for $0\nu\beta\beta$-decay was presented in \cite{phasespace}. 
The exact Dirac wave functions with finite nuclear size and electron screening 
were considered. It is believed that the calculated phase factors are not a source of uncertainty 
in the determination of the effective Majorana mass from the measured half-life.

The future experiments, CUORE (${^{130}Te}$), EXO, KamLAND-Zen (${^{136}Xe}$),
MAJORANA/GERDA (${^{76}Ge}$), SuperNEMO (${^{82}Se}$), SNO+ (${^{150}Nd}$),
and others \cite{AEE08}, with a sensitivity
\begin{equation}\label{sensitiv}
|m_{\beta\beta}|\simeq \mathrm{a~few}~10^{-2}~\mathrm{eV}
\end{equation}
will probe the IH of neutrino masses.
In the case of the normal mass hierarchy $|m_{\beta\beta}|$ is much too small in
order that $0\nu\beta\beta$-decay will be detected in experiments of the next
generation.

If the $0\nu\beta\beta$-decay will be observed, the measured half-life
$T^{0\nu-exp}_{1/2}$ with experimental
error $\sigma_{exp}$ can be converted into an ``observed
effective Majorana  mass'' $|m_{\beta\beta}|^{obs}$ and its error
$\sigma_{\beta\beta}$ as
\begin{equation}
\frac{\sigma_{\beta\beta}}{|m_{\beta\beta}|^{obs}} =
\sqrt{\frac{1}{4} \left(\frac{\sigma_{exp}}{T^{0\nu-obs}_{1/2}} \right)^2 +
\left(\frac{\sigma_{th}}{|{M'}^{0\nu}|}\right)^2 }.
\end{equation}
Here, $\sigma_{th}$ is the ``theoretical error'' of the  nuclear matrix element $|{M'}^{0\nu}|$.

In Fig. \ref{fig.2} we plot the difference $\alpha_{21}$ of the CP Majorana phases 
as function of $|m_{\beta\beta}|^{obs}$ for the inverted hierarchy 
of neutrino masses and by assuming 0\% (blue region), 15\% (red region) and 
25\% uncertainty (orange region) in $|m_{\beta\beta}|^{obs}$. The current experimental 
errors of neutrino mixing parameters and mass squared differences are taken into account. 
The lightest neutrino mass $m_0$ is assumed to be within a range from 0 to 10 meV and
one of the CP violating phases 
$\alpha_1$ (or $\alpha_2$) is taken to be arbitrary.  We see that if the considered accuracies 
are achieved it can be  possible to determine the value of the $CP$ phase difference $\alpha_{12}$.  
However, for $\sigma_{\beta\beta}/|m_{\beta\beta}|^{obs} > $ 50\% it will be  difficult, or
even impossible, to gain reliable information on the value of $\alpha_{12}$.

The uncertainty $\sigma_{th}$ in the calculated $0\nu\beta\beta$ decay
NME is a complicated and more involved problem. ${M'}^{0\nu}$ consists of the Fermi (F),
Gamow-Teller (GT) and tensor (T) parts
\cite{qrpa1,anatomy,src09}:
\begin{equation}
{M'}^{0\nu} =  \left(\frac{g^{eff}_A}{g_A}\right)^2
\left(
- \frac{M^{0\nu}_{F}}{(g^{eff}_A)^2} + M^{0\nu}_{GT} - M^{0\nu}_T
\right)~.
\label{nmep}
\end{equation}
Here, $g^{eff}_A$ is the quenched axial-vector coupling constant.
${M'}^{0\nu}$  is a function of $(g^{eff}_A)^2$, which appears in the Fermi matrix element
and also enters in the calculation of the Gamow-Teller and tensor constituents due
to a consideration of the nucleon weak-magnetism terms \cite{Sim99}.
This definition of ${M'}^{0\nu}$ allows to display
the effects of the uncertainties in $g^{eff}_A$ and to use the same phase factor $G^{0\nu}$
when calculating  the $0\nu\beta\beta$-decay rate.

The treatment of  quenching $g^{eff}_A$ is an important source of
difference between the calculated $0\nu\beta\beta$-decay NMEs \cite{anatomy,rel0n2n}.
Quenching of the axial-vector coupling was introduced to account for the fact
that the calculated strengths of the Gamow­Teller $\beta$-decay transitions
to individual final states  are significantly larger than the experimental ones.
Formally this is accomplished by replacing the true vacuum value of the
coupling constant $g_A = 1.269$  by a quenched value  $g^{eff}_A =1.0$.
It is not clear whether a similar phenomenon exists for other
multipoles besides $J = 1^+$.

Different nuclear structure methods have been used for the calculation of ${M'}^{0\nu}$,
in particular the interacting shell model (ISM) \cite{lssm,horoi10}, the quasiparticle random
phase approximation (QRPA) \cite{qrpa1,anatomy,src09},
the projected Hartree-Fock Bogoliubov approach (PHFB, PQQ2 parametrization) )
\protect\cite{phfb}, the energy density functional method (EDF) \cite{edf} and the
interacting boson model (IBM) \protect\cite{IBM09}. By assuming an unquenched
$g_A$  the ISM values of the NMEs are about a factor 2-3 smaller than the NMEs of other 
approaches (see Table 3  of \cite{VES12}). The results
of the QRPA, IBM, EDF and PHFB approaches differ by a factor less than two. Their results agree
rather well with each other  in the case of the $0\nu\beta\beta$-decay of $^{130}$Te.

A detailed study of uncertainties in the calculated $0\nu\beta\beta$-decay NMEs
was performed within  the QRPA approach \cite{qrpa1,anatomy,src09}. 
The average matrix element $\langle{M'}^{0\nu}\rangle$
(averaged over different NN potentials, choices for the single particle
space, variants of the QRPA approach) was evaluated as well as its variance $\sigma$ :
\begin{equation}
\sigma^2 = \frac{1}{N-1} \sum_{i=1}^N \left({M'}^{0\nu}_i - \langle{M'}^{0\nu}\rangle\right)^2.
\label{sigma}
\end{equation}
Further progress was achieved by performing a self-consistent calculation of the NMEs in which
the pairing and residual interactions as well as the two-nucleon short-range correlations
were derived from the same modern nucleon-nucleon potentials, namely, from the
charge-dependent Bonn potential (CD-Bonn) and the Argonne V18 potential \cite{src09}. 
The particle-particle strength of neutron-proton interaction was adjusted to 
the $2\nu\beta\beta$-decay half-life eliminating one of the main reasons for variability 
of the calculated ${M'}^{0\nu}$ within the QRPA-like methods  \cite{qrpa1}. 
We note that this procedure of fixing the particle-particle strength was also used in 
some earlier works \cite{stoica}, however, without pointing out this important 
consequence.

Recently, a further refinement of the QRPA method has been achieved 
by introducing a partial restoration of the isospin symmetry \cite{qrpawir}.
The particle-particle neutron-proton interaction was separated  into its isovector and isoscalar
parts and each were renormalized separately. The isoscalar channel of the NN interaction was fitted 
from the requirement that the calculated $2\nu\beta\beta$-decay half-life reproduces 
the experimental value. The strength of the isovector NN interaction was found to be
close to the strength of the pairing interaction following the requirement of isospin
symmetry of the particle-particle force., i.e. essentially no new parameter was introduced.

\begin{table}[!t] 
\caption{Nuclear matrix element ${M'}^{0\nu}$ for $^{76}Ge$, $^{130}Te$ and $^{136}Xe$ calculated
in QRPA with partial restoration of isospin symmetry \cite{qrpawir}.
Three different sizes of the single-particle space, two different types of NN interaction 
(CD-Bonn and Argonne) and quenched ($g_A =1.00$) or unquenched ($g_A =1.269$) 
values of the axial-vector coupling constant  are considered, i.e. 12 values are presented for each isotope.
The corresponding average matrix element $\langle{M'}^{0\nu}\rangle$ was evaluated as well as 
its variance $\sigma$ (in parentheses) following Eq. (\ref{sigma}).
$G^{0\nu}(Q_{\beta\beta},Z)$ is the phase-space factor, which values are taken from Ref. \cite{phasespace}. 
The nuclear radius $R=r_0 A^{1/3}$ with $r_0=1.2$ fm is used.
}\label{table.1}    
\centering 
\renewcommand{\arraystretch}{1.1}  
\begin{tabular}{lccccccccc}
\hline \hline 
nucleus &         &       &  & \multicolumn{3}{c}{$M'^{0\nu}$} & &  $\langle{M'}^{0\nu}\rangle$ ($\sigma$)  &    
$G^{0\nu}(Q_{\beta\beta},Z)$ \\ \cline{5-7}  
        & NN pot. & $g_A$ &  & min. s.p. space & interm. s.p. space & largest s.p. space & &   & [$y^{-1}$] \\   \hline
${^{76}Ge}$ & Argonne &  1.00  & & 3.875  & 3.701  &  3.886   & & 4.62(0.70) &  $2.36\times 10^{-15}$  \\
            &         &  1.269 & & 5.134  & 4.847  &  5.157  & &    &     \\
           & CD-Bonn &   1.00  & & 4.161  & 4.034  &  4.211  & &    &     \\
            &         &  1.269 & & 5.514  & 5.290  &  5.571  & &    &     \\
${^{130}Te}$ & Argonne &  1.00  & & 2.992  & 3.161  & 2.945  & & 3.73(0.61)  & $14.22\times 10^{-15}$  \\
            &         &  1.269 & & 3.989  & 4.229  & 3.888  & &    &     \\
           & CD-Bonn &   1.00  & & 3.317  & 3.492  & 3.297  & &    &     \\
            &         &  1.269 & & 4.438  & 4.683  & 4.373  & &    &     \\ 
${^{136}Xe}$ & Argonne &  1.00  & & 1.761  & 1.867  & 1.643  & & 2.17(0.37)   & $14.58\times 10^{-15}$  \\
            &         &  1.269 & & 2.360  & 2.509  & 2.177  & &    &     \\
           & CD-Bonn &   1.00  & & 1.963  & 2.069  & 1.847  & &    &     \\
            &         &  1.269 & & 2.639  & 2.787  & 2.460  & &    &     \\  \hline \hline  
\end{tabular}  
\end{table}    

Here, we update the calculation of the average $\langle{M'}^{0\nu}\rangle$ and  its variance
$\sigma$ for the $0\nu\beta\beta$-decay of $^{76}$Ge, $^{130}$Te and $^{136}$Xe. The recommended
half-life value of $2\nu\beta\beta$-decay of $^{76}$Ge \cite{recval} and the recently measured
half-life of the $2\nu\beta\beta$-decay of $^{136}$Xe \cite{EXO12,kamlandzen} were considered.
The calculations were performed for the CD-Bonn and Argonne potentials, 
3 different sizes of the model space \cite{qrpa1}, the unquenched or quenched value of the  
axial--vector coupling constant. 

The calculated sets of N=12 NMEs for each of the three considered isotopes are presented
in Table \ref{table.1}. The results do not depend much on the size of 
the model space and on the type of the NN interaction. For a quenched weak coupling 
constant the NMEs are significantly smaller than those for an unquenched $g_A$ mostly 
due to the factor $(1.00/1.269)=0.62$ entering in the definition of ${M'}^{0\nu}$  in 
Eq. (\ref{nmep}). The largest value of the average matrix element $\langle{M'}^{0\nu}\rangle$ 
is for $^{76}Ge$ (4.62) followed by those for $^{130}$Te (3.73) and $^{136}$Xe (2.17), 
which is about half the value. The variance $\sigma$ is about 15\% of 
the full NME $\langle{M'}^{0\nu}\rangle$. Of course, these results are only valid for the
QRPA approach and the considered averaging scheme.

It goes without saying that further progress in the calculation of the $0\nu\beta\beta$-decay
NMEs is required. Thanks to the theoretical efforts made over the last years the disagreement
among the different NMEs is now much less severe than it was some years ago. Currently the
main issue of interest is that there exists significant disagreement of the ISM results with those of
other approaches and the problem of quenching the axial-vector coupling constant.
The uncertainty associated with the calculation  of the $0\nu\beta\beta$-decay NMEs can
be reduced by suitably chosen nuclear probes. Complementary experimental information
from related processes like the $2\nu\beta\beta$-decay \cite{recval}, charge-exchange \cite{Frekers10}
and particle transfer reactions \cite{sch08}
is also very important. The differences between the results of various nuclear structure
approaches could be understood by performing an anatomy of the $0\nu\beta\beta$-decay NME
\cite{anatomy,rel0n2n,ocup09} .
The recent development in the field is very encouraging. There is reason to believe
that the uncertainty in the $0\nu\beta\beta$-decay will be further reduced.

\section{Summary}

The possible establishment of CP-violation in the lepton sector is one of the most challenging problems
of neutrino and astrophysics. Studies of $0\nu\beta\beta$-decay driven by Majorana neutrinos can lead
to insights into CP violation in this sector.
In view of recent measurement of the smallest neutrino
mixing angle $\theta_{13}$ in the Double Chooz, Daya Bay and RENO experiments we revisited the problem of determining the
Majorana CP phases by assuming the additional observation of the  $0\nu\beta\beta$-decay.

Both cases of the normal and inverted hierarchy of neutrino masses
were discussed. It was shown that in the case of the NH the determination of one of the Majorana
CP phases could 
be possible only by knowledge  of both the absolute value of the effective Majorana mass
$|m_{\beta\beta}|$ and of the lightest neutrino mass $m_1$. This task cannot be solved by 
any of the planned or prepared neutrino experiments, only within some particle physics models
which allow for a prediction of neutrino masses.
It was also found that for some values of $m_1$ the minimal value of
$|m_{\beta\beta}|$ is realized in the case of CP violation.

The case of the IH of neutrino masses offers different possibilities. Future cosmological measurements
have the potential to constrain the lightest  neutrino mass {\bf $m_0$} to values below 10 meV.
The difference $\alpha_{21}$ of Majorana phases depends very weakly on $m_0$ for these low values
and can be
determined by an accurate value for $|m_{\beta\beta}|$.
For this purpose the $0\nu\beta\beta$-decay NME needs to be evaluated with an uncertainty
of less than 30\%. This is a
formidable task, which might be achieved at some point in time due to further developments
in the fields of nuclear
structure and many-body physics also linked to a further increase of computer power.

\acknowledgments

This work is supported in part by the Deutsche Forschungsgemeinschaft
within the project "Nuclear matrix elements of Neutrino Physics and Cosmology"
FA67/40-1 and  by RFBR Grant No. 13-02-01442. F. \v S. acknowledges the support 
by the VEGA Grant agency  of the Slovak Republic under the contract No. 1/0876/12 
and by the Ministry of Education, Youth and Sports of the Czech Republic 
under contract LM2011027.


\end{document}